\documentclass[aps,prb,twocolumn,showpacs,20pt]{revtex4}
\usepackage{epsfig}
\begin{document}

\def \cT {{\cal T}}
\def \cI {{\cal I}}
\def \cf {{\cal f}}
\def \cG {{\cal G}}
\def \cD {{\cal D}}
\def \cU {{\cal U}}
\def \cV {{\cal V}}
\def \cF {{\cal F}}
\def \cT {{\cal T}}
\def \cH {{\cal H}}
\def \cA {{\cal A}}
\def \cL {{\cal L}}
\def \cR {{\cal R}}
\def \cN {{\cal N}}
\def \cC {{\cal C}}
\def \cS {{\cal S}}
\def \cP {{\cal P}}
\def \cE {{\cal E}}
\def \cM {{\cal M}}

\title{Interference in transport through double barriers in interacting quantum wires}

\author{Shaoqin Wang$^1$}
\author{Liling Zhou$^{2,3}$}
\author{Zhao Yang Zeng$^1$}
\email{zyzeng@jxnu.edu.cn}

 \affiliation {$^1$Department of Physics,
Jiangxi Normal University, Nanchang 330022, China\\
$^2$Department of Physics,
Jiujiang University, Jiujiang 3320052, China\\
$^3$State Key Laboratory for Superlattices and
Microstructures, Institute of Semiconductors, Chinese Academy of
Sciences, Beijing 100083, China}

\begin{abstract}
We investigate interference effects of the backscattering current through a double-barrier structure in an interacting quantum wire  attached to  noninteracting leads. Depending on the interaction strength and the location of the barriers, the backscattering current exhibits different oscillation and scaling characteristics with the applied voltage in the strong and weak interaction cases. However, in both cases, the oscillation behaviors of the backscattering current are mainly determined by the quantum mechanical interference due to the existence of the double barriers.

\end{abstract}
 \maketitle

\section{introduction}
  As a fundamental many-body physical model, one-dimensional($1$D) interacting electron systems are an everlasting research topic. Unlike its high-dimensional counterparts, which are well understood from the quasiparticle picture  in the Fermi liquid theory,\cite{Nozieres}  $1$D interacting electron systems can be described  by bosonic gapless collective excitations of the fermion density fluctuations within the framework of the Tomonaga-Luttinger liquid theory.\cite{Tomonaga,Luttinger,Haldane}

  Tomonaga-Luttinger liquid behaviors in $1$D systems have been revealed in measuring the transport properties of cleaved-edge overgrowth quantum wires,\cite{quantum wire} quantum Hall systems,\cite{quantum hall} and  single-wall carbon nanotubes.\cite{nanotube} It opens up the possibilities to test some theoretical predictions even in  simplified pure physical models, and  triggers intensive attention to the physics of the Tomonaga-Luttinger liquids.\cite{Giamarchi}

  In mesoscopic Fermi-liquid systems, electronic transport is modeled by the transmission of an incident electron through a potential barrier, and the conductance of the system is directly related to the transmission probability.\cite{Landauer} However, potential barriers play
  a counterintuitive role in the Tomonaga-Luttinger liquids. In a seminal paper, Kane and Fisher\cite{Kane} have shown that the barrier is irrelevant for attractive electron-electron interactions, and cuts the system into two pieces for repulsive interactions. In  transport measurements, Tomonaga-luttinger liquid quantum wires of finite length need to be connected to leads acting as electron reservoirs. If the reservoirs are modeled as $1$D noninteracting systems, then the model of inhomogeneous Tomonaga-luttinger liquids with different interaction strengths in different parts  is appropriate to investigate the transport properties of the Tomonaga-luttinger liquids.\cite{Maslov,Safi}  It was argued that the dc conductance is not renormalized by the interactions.\cite{Maslov,Safi} An interesting phenomenon of Andreev-like reflections has been shown at the interfaces between the interacting wire and the noninteracting leads due to the mismatch of the interaction strengths.\cite{Safi} The presence of an impurity results in interference of the bosonic excitations which can be modulated by an applied
  bias voltage, and thus leads to characteristic oscillations of the backscattering current as a function of the dc voltage.\cite{Dolcini}
  Feldman et al.\cite{Feldman} have showed that backscattering off a weak dynamic impurity would enhance the current. Pe\c{c}a et al.\cite{Peca} and Recher et al.\cite{Recher} have demonstrated interesting Fabry-Perot interference patterns of the non-linear conductance as a function of the bias voltage in carbon nanotubes attached to metallic reservoirs, where backscattering processes mainly occur at the two metal-contact-nanotube interfaces.

  In this paper, we investigate interference effects of electron tunneling through double barriers in an interacting quantum wire.  We assume that the $1$D interacting electron system is  adiabatically attached to two $1$D noninteracting electron systems, which act as electron reservoirs. Such an assumption excludes the possibility of backscattering at the interfaces between the interacting and noninteracting segments. Backscattering events take place at the positions of the double  barriers. Due to the existence of the Andreev-like reflections at the interfaces, electron tunneling through the double barriers in a Tomonaga-Luttinger-liquid wire is more complicated and thus expected to be more interesting. As we show later, there exists a contribution from the quantum mechanical interference of bosonic excitations  to  the backscattering current, besides the contributions from backscattering off the separate barriers.  It is the quantum mechanical interference that dominates the oscillatory characteristics of the backscattering current as a function of the applied bias voltage. The aim of this work is to make clear  how the oscillation pattern of the backscattering current in an interacting quantum wire with double barriers is modulated by the electron interaction strength and the arrangement of the double barriers.

\section{mode and formulation}

We consider an interacting quantum wire of length $L$ with double  point-like barriers, which is
at its ends connected  adiabatically to two semi-infinite noninteracting  quantum wires acting as
electron reservoirs with electrochemical potentials $\mu_L$ and $\mu_R$ .
The model Hamiltonian can be written as
\begin{eqnarray}
\label{The total Hamitonian}
H&=&H_W+H_B, \nonumber \\
H_W&=&\int^{+\infty}_{-\infty} dx\Big\{ \psi^\dagger(x)\big[\frac{-\hbar^2}{2m}\frac{d^2}{dx^2}+\mu(x)\big]\psi(x)\nonumber\\
&&~~~~~~~~+U(x)\psi^{\dagger 2}(x)\psi^2(x)\Big\}, \nonumber \\
H_B&=& \int dx \psi^\dagger(x)\lambda\big[\delta(x-x_1)+\lambda\delta(x-x_2)\big]\psi(x),  \nonumber                                                \end{eqnarray}
where $H_W$  describes both the interacting wire and the interacting leads with varied electron-electron interaction strengths, i.e.   $U(x)=U$
 if $-L/2\leq x \leq L/2$ and   $U(x)=0$  elsewhere, $H_B$ represents the Hamiltonian for the point-like barriers, and the function
$\mu(x)=\mu_L[1-\theta(x+L/2)]+\mu_R\theta(x-L/2)$ denotes the externally tunable electrochemical potential.

   Since what we are interested in is the low-temperature transport properties, it is convenient to reformulate the problem in the framework of standard bosonization.\cite{Bosonization}  First linearize the energy spectra about the Fermi points $\pm k_F$, and introduce two species of electron operators $\psi_\alpha(x)(\alpha=R/L=\pm$) to describe the right-moving and left-moving fermions, then  the excitations of the fermion system can be described by a bosonic  field $\vartheta(x)$. In this way the Hamiltonian for the fermion system can be recast in the following bosonic form

\begin{eqnarray}
H&=&H_W+H_B,  \\
\label{The H_W}
H_W &=& \frac \hbar 2\int dx\frac{1}{g(x)}\Big[\frac{1}{v(x)}
(\partial_t\vartheta)^2+v(x)(\partial_x\vartheta)^2\Big] \nonumber\\
&&~~~~-\frac{e}{\sqrt{\pi}}\int^{+\infty}_{-\infty}dx E(x,t)\vartheta(x,t),\\
\label{TheH_B}
H_B &=&\lambda\cos(\sqrt{4\pi}\vartheta(x_1,t)+2k_F x_1)\nonumber\\
&&~~~~+\lambda\cos(\sqrt{4\pi}\vartheta(x_2,t)+2k_F x_2),
\end{eqnarray}
where $g(x)=[1+U(x)/ \pi \hbar v_F]^{-1/2}$ is the interaction parameter, $-eE(x,t)=\partial_x\mu=-\mu_L\delta(x+L/2)+\mu_R\delta(x-L/2)$, and $v(x)=v_F/g(x)$ is the charge density wave velocity.

In terms of the Bosonic field $\vartheta$, the current operator can be written as
\begin{eqnarray}
\label{current operator}
\hat{I}(x,t)=\frac{e}{\sqrt{\pi}}\partial_t\vartheta(x,t).
\end{eqnarray}

The current-bias relations of the nonequilibrium system is of interest, it is appropriate  to adopt  the Keldysh formalism.  Following the path integral technical procedures developed in Ref. $13$  for a static impurity, the average current is obtained
\begin{eqnarray}
\label{total current}
\langle I(x,t)\rangle=\langle I_0- I_{BS}(x,t)\rangle.
\end{eqnarray}
In Eq. (\ref{total current}), $\langle I_0 \rangle=e^2 V/h$ is the background current in the absence of barriers, and  $\langle I_{BS}(x,t)\rangle=-e\langle \frac{d :\psi_R^{\dagger}(x,t)\psi_R(x,t):}{dt}\rangle$ is the backscattering current. The average backscattering current is  the sum of three contributions to leading order in $\lambda$ (see Appendix)

\begin{eqnarray}
I_{BS}=&&I_{BS}^1+I_{BS}^2+I_{BS}^{12}\nonumber \\
=&&\frac{e\lambda^2}{4\hbar^2}\int^{+\infty}_{-\infty}
dt\sin\omega_0t \Big[e^{4\pi C_0(x_1,t;x_1,0)}+\nonumber \\
&&~~~~~~~ e^{4\pi
C_0(x_2,t;x_2,0)}
 +2e^{4\pi
C_0(x_1,t;x_2,0)}\Big],
\end{eqnarray}
where $\omega_0=eV/\hbar$, $C_0(x,t;x',0)=\Big\langle\vartheta(x,t)\vartheta(x',0)
-[\vartheta^2(x,t)+\vartheta^2(x',0)]/2\Big \rangle_0$ are the
correlation functions of the bosonic field $\vartheta(x,t)$ in the
clean wire limit. These correlation functions can be obtained by expanding the Bosonic field on the basis of the eigenfunctions, which satisfy a specific inhomogeneous equation.\cite{Dolcini}  It is evident that, besides the independent contributions from backscattering by the separate  barriers, there exists a quantum mechanical interference contribution to the backscattering
current due to the coexistence of double barriers. It is just such an interference term  that gives rise to some interesting transport characteristics in interacting
quantum wires with double barriers.

   At zero temperature, the correlation
functions yield a simplified expression
\begin{eqnarray}
&&C_0(x_i , t;x_j,0)= \nonumber \\
&&-\frac{g}{4\pi}\Big\{\sum_{m\in
even}\gamma^{|m|}\ln\frac{(\alpha+i\tau)^2+(m+|\xi_i-\xi_j|)^2}{\alpha^2+m^2}
\nonumber \\
&&+\sum_{m\in
odd}\gamma^{|m|}\Big[\ln\Big(\frac{(\alpha+i\tau)^2+(m-\xi_i-\xi_j)^2}{\alpha^2+(m-\xi_i-\xi_j)^2}\Big)\nonumber\\
&&+\frac{1}{2}\ln\frac{[\alpha^2+(m+ \xi_i+\xi_j                               )^2]^2}{[\alpha^2+(2\xi_i+m)^2][\alpha^2+(2\xi_j+m)^2]}
\Big]\Big\},\nonumber
\end{eqnarray}
where $\xi_i=x_i/L$,  $\alpha=\omega_L/\omega_c$ is the
dimensionless cutoff, with  $\omega_L=V_F/gL$ being the inverse of the traversal time of the charge density wave and
$\omega_c$ being the high-energy cutoff frequency,
$\gamma=(1-g)/(1+g)$ is the Andreev-like reflection coefficient at the interfaces.

In terms of the dimensionless variables
$u=eV/\hbar\omega_L$, $\tau=\omega_Lt$, the backscattering current is rewritten into the following
simple form  in unit of $e(\lambda\omega_L^g/\omega\omega_c^g)^2/\hbar^2\omega_L$
\begin{eqnarray}
\label{bc}
I_{BS}=\frac{i\alpha^{-2g}}{2}\int^{+\infty}_{-\infty} d\tau\sin
u\tau(A_1+A_2+2A_{12}),
\end{eqnarray}
with
\begin{eqnarray}
A_{1/2}&=&\prod_{m\in
even}\Big[\frac{(\alpha+i\tau)^2+m^2}{\alpha^2+m^2}\Big]^{-g\gamma^{|m|}}
\times
\nonumber \\
&&~~~ \prod_{m\in
odd}\Big[\frac{(\alpha+i\tau)^2+(m-2|\xi_{1/2}|)^2}{\alpha^2+(m-2|\xi_{1/2}|)^2}
\Big]^{-g\gamma^{|m|}},\nonumber
\end{eqnarray}
\begin{eqnarray}
A_{12}&=&\prod_{m\in
even}\Big[\frac{(\alpha+i\tau)^2+(m+|\xi_1-\xi_2|)^2}{\alpha^2+m^2}\Big]^{-g\gamma^{|m|}}
\times\nonumber \\
&&~~\prod_{m\in
odd}\Big[\frac{(\alpha+i\tau)^2+(m-\xi_1-\xi_2)^2}{\alpha^2+(m-\xi_1-\xi_2)^2}\Big]^{-g\gamma^{|m|}}\nonumber\\
&&\times\Big\{\frac{[\alpha^2+(m+\xi_1+\xi_2)^2]^2}{[\alpha^2+(m+2\xi_1)^2][\alpha^2+(m+2\xi_2)^2]}
\Big\}^{-\frac{1}{2}g\gamma^{|m|}}.\nonumber
\end{eqnarray}

The above expressions for the backscattering current $I_{BS}$ allow us to  evaluate it for arbitrary values of
the barrier positions ($x_1,x_2$), the interaction strength $g$, and the applied voltage
$V$. It is noticed that the backscattering current contributed  from an individual barrier is  dependent on the relative distance of the barrier to the wire center($|\xi_{1/2}|$), while that from the quantum mechanical interference term is dependent on the location details of the double barriers($|\xi_1-\xi_2|$ and $\xi_1+\xi_2$). It seems impossible to obtain an explicit expression for the backscattering current, we resort to numerical
calculations  to investigate the backscattering current in some typical cases.

\section{results and discussions}

First, we investigate the dependence of the backscattering current on the interaction strength. It is well known that the interaction
strength is characterized by the interaction parameter $g=[1+U/ \pi \hbar v_F]^{-1/2}$. We have $0<g < 1$ for repulsive interactions and  $g=1$ for noninteracting limit. It should be noted that the smaller the value of $g$, the stronger the interaction.  In Fig. $1$, we present numerical results of  $I_{BS}$ as a function of the applied voltage $u$ for different typical interaction parameters $g=0.25,0.5,0.75$,  as the barriers locate symmetrically near the ends of the wire with $\xi_{1/2}=\pm 0.4$. Since the barriers are symmetrically located,  the backscattering currents arising from  different barriers are the same. It can be observed that, as the interaction decreases, the oscillation of the incoherent addition of the backscattering currents contributed from the two barriers disappears gradually, while that contributed from the quantum mechanical interference persists and is more prominent.  The oscillatory behavior of the backscattering current is mainly determined by the  quantum mechanical interference of the bosonic waves backscattering off the double barriers. We attribute the less pronounced oscillation of the backscattering current in the strong interaction case to the  suppression of the quantum mechanical interference by the electron-electron interaction. As shown by Dolcini et al. \cite{Dolcini}, the oscillation of the backscattering current in the single barrier case arises from a combined effect of the barrier, the finite length and the interaction in the wire. The phase shift of bosonic excitations traveling between the barrier and the interfaces is responsible for such an oscillation and can be modulated by an applied bias voltage. In our case with double barriers, competition between the single-barrier interference and the double-barrier quantum mechanical interference results in an interesting oscillatory behavior of the backscattering current. In the cases of a given barrier location, the oscillation period of the backscattering current contributed from  the quantum mechanical interference  is irrespective of the interaction strength, while the oscillation period from the incoherent addition of the backscattering currents off different barriers is increased as the interaction parameter is decreased, and eventually becomes infinite in the noninteracting limit,  which can be found in Fig. $1$.

\begin{figure}\epsfig{file=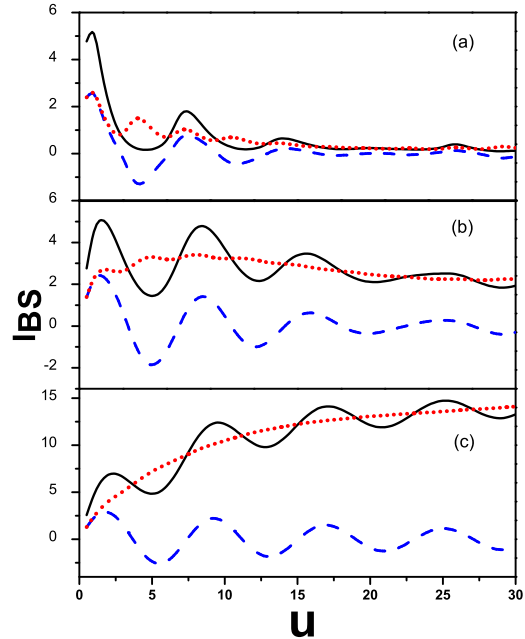, width=8cm}
\caption{Backscattering current $I_{BS}$ (in unit of
$e(\lambda\omega_L^g/\omega_c^g)^2/\hbar^2\omega_L$) as a function
of $u=eV/\hbar\omega_L$ with the double barrier location $\xi_{1/2}=x_{1/2}/L=\pm0.4$ for different
interaction parameters (a) $g=0.25$, (b) $g=0.5$ and (c) $g=0.75$, respectively. The solid line refers to the
total backscattering current $I_{BS}$, the dot line to the
incoherent addition of independent contributions from  different barriers $I_{BS}^1+I_{BS}^2$ and
the dash line to the quantum mechanical interference term $I_{BS}^{12}$.}
\end{figure}

Then we consider two special cases for strong interaction ($g=0.25$) where the double barriers locate symmetrically near the midpoint of the interacting wire ($\xi_{1/2}=\pm 0.01$)  and  right at the ends of the wire ($\xi_{1/2}=\pm 0.5$).  It can be found from Fig. $2$ that in both cases the backscattering current oscillates in a more pronounced way,  compared to the case where the double barriers are located symmetrically near the interfaces ($\xi_{1/2}=\pm 0.4$, Fig. $1$ (a)). In Fig. $2$,  we also observe that, the oscillation frequency of the incoherent addition of the  backscattering currents off  the separate  barriers is twice that due to the quantum mechanical interference when the double barriers are located at the ends of the wire, while the two frequencies are about the same when the double barriers are located near the midpoint of the wire. The reason is that the round-trip ballistic time for bosonic excitations propagating between a barrier and an interface in the former case is twice that in the later case.

\begin{figure}\epsfig{file=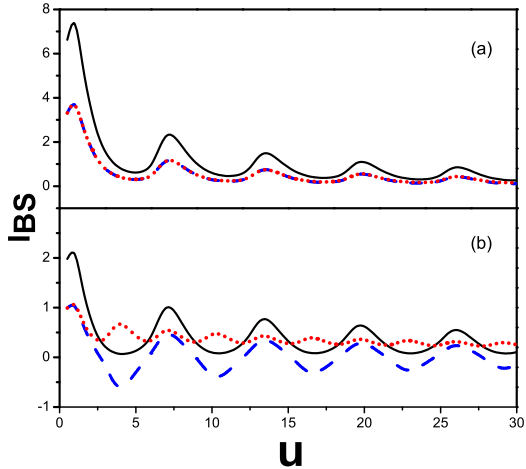, width=8cm}
\caption{Backscattering current $I_{BS}$ (in unit of
$e(\lambda\omega_L^g/\omega_c^g)^2/\hbar^2\omega_L$) as a function
of $u=eV/\hbar\omega_L$ as $g=0.25$ with different double barrier locations
(a) $\xi_{1/2}=x_{1/2}/L=\pm0.01$ and (b) $\xi_{1/2}=x_{1/2}/L=\pm0.5$.  The solid line refers to the
total backscattering current $I_{BS}$, the dot line to the
incoherent addition of independent contributions from  different barriers $I_{BS}^1+I_{BS}^2$ and
the dash line to the quantum mechanical interference term $I_{BS}^{12}$.}
\end{figure}

Up to the present we have analyzed the current-voltage characteristics in cases where the two barriers are symmetrically located. As the barriers are located asymmetrically,
it is expected that such an asymmetry has trivial effects on the current-voltage characteristics of weakly interacting wires, but influences significantly the transport properties of strongly interacting wires. In Fig. $3$, we provide the results of  the backscattering current of the interacting wire with symmetric  and asymmetric locations of the barriers in both weak interaction($g=0.75$) and strong interaction($g=0.25$) cases. We find that in the weak interaction case, the backscattering currents are about the same for symmetric and asymmetric barrier locations(Figs. $3$(a) and $3$(b)), as long as the spacing between the double barriers is fixed. However, such a scenario is changed in the strong interaction case, where the current oscillation strongly depend on the symmetry of the barrier locations (Figs. $3$(c) and $3$(d)). It is interesting to note that, the period of current oscillation depends strongly on the spacing between the double barriers in the weak interaction limit, i.e., the bigger the barrier spacing, the smaller the oscillation period. This phenomenon can not be observed in the strong interaction limit. We attribute the dependence of the current oscillation on the barrier spacing in the weak interaction limit to the resonant tunneling through a double-barrier structure in one-dimensional noninteracting electronic systems.\cite{Resonant tunnelling} In the case of strong interactions, different locations of the barriers lead to different interference patterns for bosonic plamonic excitations traveling between a barrier and the interfaces,  and then results in different oscillation behaviors of the backscattering currents contributed from the separate barriers.

Finally, we would like to point out that, depending on the interaction strength, the backscattering current exhibits different scaling rules with the applied bias voltage in a finite-length Tomonaga-Luttinger-liquid wire with double barriers, as can be found from Figs. $1,2,3$. This observation deserves further investigation.

\begin{figure}\epsfig{file=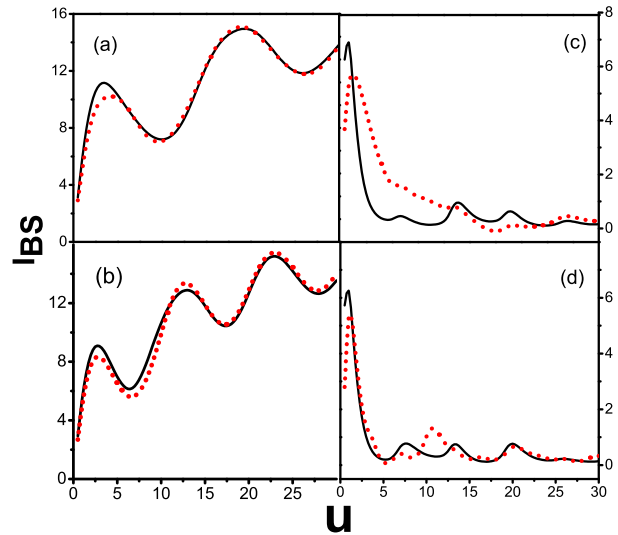, width=9cm }
\caption{Backscattering current $I_{BS}$ (in unit of
$e(\lambda\omega_L^g/\omega_c^g)^2/\hbar^2\omega_L$) as a function
of $u=eV/\hbar\omega_L$ with different double barrier locations and for different
interaction parameters (a)  $g=0.75$, $\xi_{1/2}=x_{1/2}/L=\pm0.2$ (solid), $\xi_{1/2}=x_{1/2}/L=0,0.4$ (dotted) (b) $g=0.75$,
 $\xi_{1/2}=x_{1/2}/L=\pm0.3$ (solid), $\xi_{1/2}=x_{1/2}/L=-0.15,0.45$ (dotted),  (c)  $g=0.25$, $\xi_{1/2}=x_{1/2}/L=\pm0.2$ (solid), $\xi_{1/2}=x_{1/2}/L=0,0.4$ (dotted), and (d) $g=0.25$, $\xi_{1/2}=x_{1/2}/L=\pm0.3$ (solid), $\xi_{1/2}=x_{1/2}/L=-0.15,0.45$ (dotted).}
\end{figure}

\section{conclusion}
In summary, we have investigated interference effects in electron transport through a double-barrier structure in an interacting quantum wire, which is ideally attached to two noninteracting leads. We have found that the oscillation of the backscattering current with the applied voltage is mainly determined by the
quantum mechanical interference due to the coexistence of the double barriers. It is contrast to the single barrier case, where the current oscillation
is strongly dependent on the interference of the Andreev-like reflected plasmonic excitations propagating between the single barrier and the interfaces. Depending on the interaction strength and the locations of the barriers, the competition between these two kinds of interferences results in different oscillation and scaling behaviors of the backscattering current with the bias voltage.

\acknowledgments

This work is supported by the NSFC under Grant No. 10404010, the
Project-sponsored by SRF for ROCS, SEM and the Excellent talent
fund of Jiangxi Normal University.

\appendix
\section{Outline of derivation of backscattering current expression}

 In this appendix,we outline the main formulae to calculate the backscattering current based on the
Keldysh functional approach following Ref. $13$.

The generating functional takes the form
\begin{eqnarray}
Z[J]=\frac{1}{\mathcal {N}_\mathcal {Z}}\int D\vartheta
\exp\Big[\frac{i}{\hbar}\mathcal {S} +\frac{i}{\sqrt{2}}\int
d\textbf{x} J{(\textbf{x})}\vartheta{(\textbf{x})}\Big],
\end{eqnarray}
where the action functional of the system can be written in terms of
the boson field $\vartheta(x,t)$ as
\begin{eqnarray}
\mathcal {S}=&&\hbar\int dx \int dt\frac {1}{2g(x)}
\Big[\frac{1}{v(x)}
(\partial_t\vartheta)^2-v(x)(\partial_x\vartheta)^2\Big]\nonumber\\
&&+\frac{e}{\sqrt{2}}\int dx \int dt E(x,t)\vartheta(x,t)-\int dt
H_B.
\end{eqnarray}

After introduce the standard keldysh time contour and
denote by $\vartheta^+$ and $\vartheta^-$ the complex fields on the upper
and lower time branches of the Keldysh contour, and define four Green's functions averaged with respect to the free
Hamiltonian $C_{0}^{\eta\eta'}(\textbf{r}';\textbf{r}")=
\langle\vartheta^\eta(\textbf{r}')\vartheta^{\eta'}(\textbf{r}")\rangle_0$,
the generating functional is rewritten as
\begin{eqnarray}
&&Z[J]=\frac{1}{\mathcal {N}_\mathcal {Z}}\int D\vartheta^\pm \exp\Big\{-\frac{1}{2}\int
d\textbf{r}' d\textbf{r}''\sum_{\eta~\eta'=\pm}\vartheta^\eta(\textbf{r}')\nonumber\\
&&(C^{-1})^{\eta\eta'}(\textbf{r}';\textbf{r}'')\vartheta^{\eta'}
(\textbf{r}'')\Big\}\exp\Big\{\sum_{\eta=\pm}\Big(\frac{i\eta e}{\hbar\sqrt{\pi}}\int d\textbf{r}'E(\textbf{r}') \nonumber\\
&&\vartheta^\eta(\textbf{r}')-\frac{i\eta}{\hbar}\int_{-\infty}^{+\infty} dt' H_B[\vartheta^\eta]
+\frac{i}{\sqrt{2}}\int d\textbf{x}J{(\textbf{x})}\vartheta^\eta(\textbf{x}))\Big)\Big\},\nonumber
\end{eqnarray}
where $C^{-1}(\textbf{r};\textbf{r}')$ is the inverse of of a $2\times 2$ matrix formed from the above four Green's functions.

Define the following matrices
\begin{eqnarray}
&&\mbox{\boldmath$\vartheta$}=
\left( {\begin{array}{*{20}c}
   {\vartheta^+(r) } \\
   {\vartheta^-(r) } \\
\end{array}} \right) , \nonumber \\
&&\textbf{J}=\left( {\begin{array}{*{20}c}
{\frac{e}{\hbar}\sqrt{\frac{2}{\pi}}E(r)}\\
{J(r)}\\
\end{array}} \right) ,\nonumber\\
&&\textbf{Q}=\frac{1}{\sqrt{2}} \left( {\begin{array}{*{20}c}
{1}&&{-1}\\
{1}&&{1}\\
\end{array}} \right)\delta(r-r'),
\end{eqnarray}
and
\begin{eqnarray}
\textbf{C}_0=\left( {\begin{array}{*{20}c}
{C^{++}_{0}(r,r')}&&{C^{+-}_{0}(r,r')}\\
{C^{-+}_{0}(r,r')}&&{C^{--}_{0}(r,r')}\\
\end{array}} \right) ,
\end{eqnarray}
the generating functional is reduced to the following simplified form
\begin{eqnarray}
Z[J]&=&\frac{1}{\mathcal {N}_\mathcal
{Z}}\int D\mbox{\boldmath$\vartheta$}e^{(-\mbox{\boldmath$\vartheta$}^T\textbf{C}^{-1}_0\mbox{\boldmath$\vartheta$}
+2i\textbf{J}^T\textbf{Q}\mbox{\boldmath$\vartheta$}) /2}\nonumber\\
&&~~~~~~\times\exp\Big\{-\frac{i}{\hbar}\sum_{\eta=\pm}\eta\int dt'
H_B[\vartheta^\eta]\Big\}.
\end{eqnarray}

After Shift the fields $\mbox{\boldmath$\vartheta$}\rightarrow\mbox{\boldmath$\vartheta$}+\textbf{A}_J,\textbf{A}_J=i\textbf{C}_0\textbf{Q}^T\textbf{J}$,
one obtains a factorized form of the generating functional
\begin{eqnarray}
Z[J]&&=e^{-\textbf{J}^T\textbf{Q}\textbf{C}_0\textbf{Q}^T\textbf{J}/2}\nonumber\\
&&\times\Big\langle\exp(-\frac{i}{\hbar}\sum_\eta\eta\int
H_B[\vartheta^\eta+A^\eta_J]dt')\Big\rangle_0.
\end{eqnarray}

From the expression of current operator Eq. (\ref{current operator}), we have
\begin{eqnarray}
\langle I(x,t) \rangle&=&\frac{e}{\sqrt{\pi}}\partial_t \langle\vartheta(x,t)\rangle\nonumber \\
&=&\frac{e}{\sqrt{\pi}}\partial_t \big\langle \frac{-i}{\sqrt{2}}\frac{\delta Z[J]}{\delta J(\textbf{x})}\big |_{J=0}\big\rangle\nonumber\\
&=&\frac{e}{\sqrt{\pi}}\partial_t \Big[ \frac{ie}{\hbar \sqrt{\pi}}\int d\textbf{r}' C^R_0(\textbf{x};\textbf{r}')E(\textbf{r}')-\frac{1}{\sqrt 2\hbar}\nonumber\\
&&\big \langle \sum_{\eta=\pm}\int dt' \frac{\delta H_B[\vartheta^\eta+A^\eta_J]]}{\delta \vartheta^\eta}\frac{\delta A^\eta_J(\textbf{r}')}{\delta J(\textbf{x})}\big \rangle_0 \Big]\nonumber\\
&=&e^2V/h-I_{BS},
\end{eqnarray}
where $C_0^R(\textbf{r};\textbf{r}')=\theta(t-t')\langle [\vartheta(\textbf{r}),\vartheta(\textbf{r}')]\rangle_0$ is the retarded Green's function, and the backscattering current takes the following form

\begin{eqnarray}
\label{bcc}
 I_{BS}&&=-\sum_{i=1,2}\frac{\hbar \sqrt{\pi}}{e^2}\int dt'
\sigma_0(\textbf{x};\textbf{r}_i')
\langle j^+_B(\textbf{r}_i')\rangle_\rightarrow.
\end{eqnarray}
Here $\sigma_0(\textbf{r};\textbf{r}')
=2ie^2\partial_tC^R_0(\textbf{r};\textbf{r}')/h$ is the local conductivity of clean wire, and
the backscattering current operator $j_B^\eta(\textbf{x})=-\frac{e}{\hbar}\frac{\delta
H_B[\vartheta^\eta+A_0^\eta]}{\delta\vartheta(\textbf{x})}$ with $A_0(\textbf{r})=\frac{ie}{\hbar \sqrt{\pi}}\int d\textbf{x}'C^R_0(\textbf{r};\textbf{x}')E(\textbf{x}')$,
$\langle\cdot\cdot\cdot\rangle_\rightarrow$ denotes an
average along the Keldysh contour with respect to the shifted
Hamiltonian
$H_\rightarrow=H_0[\vartheta]+H_B[\vartheta+A_0]$.

Some algebras after substitution of the expression of  $H_B$  into Eq. (\ref{bcc})  finally yield
\begin{eqnarray}
I_{BS}=&&-\frac{2\pi\lambda}{e}\sum_{i=1,2}\int^{+\infty}_{-\infty}dt'\sigma_0( x,t;x_i,t')
\nonumber \\ &&\times\Big\langle\exp\Big(-\frac{i}{\hbar}
\sum_{\eta=\pm}\eta\int^{+\infty}_{-\infty}dt''H_B[\vartheta^\eta+\omega_0t'']\Big)\nonumber \\
&&~~~~\sin[\sqrt{4\pi}\vartheta^+(x_i,t')+2k_Fx_i+\omega_0t']\Big\rangle_0.
\end{eqnarray}

\end{document}